\documentclass[aps,showpacs,superscriptaddress]{revtex4}
\usepackage[T1]{fontenc}
\usepackage{amsmath}
\usepackage{amsfonts}
\usepackage{amssymb} 
\usepackage{graphics}

\usepackage{etoolbox}

\newcommand{\mysfig}[2]
{
\begin{figure}
\vskip 0.2cm
\begin{tabular}{@{}c@{}}
\resizebox{0.35\textwidth}{!}{
\includegraphics{./#1.eps}
}
\\
\begin{minipage}[c]{0.98\textwidth}
\caption{\footnotesize #2 \label{#1} }
\end{minipage}
\end{tabular}
\vskip 0.2cm
\end{figure}
}
\newcommand{\mysfigbig}[2]
{
\begin{figure}
\vskip 0.2cm
\begin{tabular}{@{}c@{}}
\resizebox{0.95\textwidth}{!}{
\includegraphics{./#1.eps}
}
\\
\begin{minipage}[c]{0.98\textwidth}
\caption{\footnotesize #2 \label{#1} }
\end{minipage}
\end{tabular}
\vskip 0.2cm
\end{figure}
}
\newcommand{\myfig}[4]
{
\begin{figure}
\vskip 0.2cm
\begin{tabular}{@{}cc@{}}
\resizebox{0.35\textwidth}{!}{
\includegraphics{./#1.eps}
}
&
\resizebox{0.35\textwidth}{!}{
\includegraphics{./#3.eps}
}
\\
\begin{minipage}[c]{0.49\textwidth}
\caption{\footnotesize #2 \label{#1} }
\end{minipage}
&
\begin{minipage}[c]{0.49\textwidth}
\caption{\footnotesize #4 \label{#3} }
\end{minipage}
\end{tabular}
\vskip 0.2cm
\end{figure}
}
\def\mysixfig#1#2#3#4#5#6#7#8#9{%
\def\ArgI{#1}%
\def\ArgII{#2}%
\def\ArgIII{#3}%
\def\ArgIV{#4}%
\def\ArgV{#5}%
\def\ArgVI{#6}%
\def\ArgVII{#7}%
\def\ArgVIII{#8}%
\def\ArgIX{#9}%
\mysixfigcont
}
\def\mysixfigcont#1#2#3{
\begin{figure}
\vskip 0.2cm
\begin{tabular}{@{}cc@{}}
\resizebox{0.35\textwidth}{!}{\includegraphics{./\ArgI.eps}}
&
\resizebox{0.35\textwidth}{!}{\includegraphics{./\ArgVII.eps}}
\\
\begin{minipage}[c]{0.49\textwidth}
\caption{\footnotesize \ArgII \label{\ArgI} }
\end{minipage}
&
\begin{minipage}[c]{0.49\textwidth}
\caption{\footnotesize \ArgVIII \label{\ArgVII} }
\end{minipage}
\\
\resizebox{0.35\textwidth}{!}{\includegraphics{./\ArgIII.eps}}
&
\resizebox{0.35\textwidth}{!}{\includegraphics{./\ArgIX.eps}}
\\
\begin{minipage}[c]{0.49\textwidth}
\caption{\footnotesize \ArgIV \label{\ArgIII} }
\end{minipage}
&
\begin{minipage}[c]{0.49\textwidth}
\caption{\footnotesize #1 \label{\ArgIX} }
\end{minipage}
\\
\resizebox{0.35\textwidth}{!}{\includegraphics{./\ArgV.eps}}
&
\resizebox{0.35\textwidth}{!}{\includegraphics{./#2.eps}}
\\
\begin{minipage}[c]{0.49\textwidth}
\caption{\footnotesize \ArgVI \label{\ArgV} }
\end{minipage}
&
\begin{minipage}[c]{0.49\textwidth}
\caption{\footnotesize #3 \label{#2} }
\end{minipage}
\end{tabular}
\vskip 0.2cm
\end{figure}
}

\newcommand{\mytab}[4]{
\begin{table}[!hbt]
\begin{flushright} Table $\;$ #1 \end{flushright}
\begin{center} \textsl{\small{#3}}\end{center}
\begin{center}
\begin{tabular}{#2}
 #4
 \end{tabular}
 \end{center}
\end{table}
 }

\begin{document}
\title{Analysis of correlations of dwell-times of adjacent kinetic states in the activity of the cold and menthol receptor TRPM8}
\author{Ogloblya O.V.}
\email{olexandr.ogloblya@gmail.com}
\affiliation{Taras Shevchenko National University, 64/13 Volodymyrska St., Kyiv 01601, Ukraine }
\author{Moroz O.F.}
\affiliation{National University «Kyiv aviation institute», 1 Liubomyra Huzara Avenue, Kyiv 03058, Ukraine}
\author{Zholos A.V.}
\affiliation{Taras Shevchenko National University, 64/13 Volodymyrska St., Kyiv 01601, Ukraine }

\date{\today}

\begin{abstract}
Temperature-sensitive transient receptor potential (TRP) channels play a significant role in intercellular signalling  in response to membrane depolarisation and caclium influx. TRPM8 ion channels have been investigated as the main cold receptors (neurosensors), but they can also be activated by voltage, $Ca^{2+}$ store depletion, and certain lipids and other ligands that induce cold sensation (such as menthol and icilin). Despite recent progress in studying these channels at the single-channel level, the mechanisms and time course of their activation remain underexplored. We propose an approach to single-ion channel analysis that elucidates the presence of correlations between dwell-times of adjacent open and closed states. One of the major benefits of this approach is that it can serve as an express test of channel activity properties owing to its simplicity and fast numerical realization. The idea of our analysis is to study the difference between the distribution function derived from the given trace (simulated or experimentally recorded) and the predicted distributions of simple models without correlation. In this way, we compared the distribution function of the difference between the durations of the adjacent open and closed states of the channel, as well as that of a similar sequence of open-closed-open-closed states.
\end{abstract}


\maketitle

\section{Introduction}

Transient receptor potential (TRP) channels play an important role in cell signalling by responding to various environmental physical 
and chemical factors and translating these responses into changes in cell membrane potential. 
The latter is achieved via sodium and calcium influx, thereby modulating intracellular calcium levels~\cite{zhang}. 
TRPM8 cation channel, beginning from its identification as cold and menthol receptor independently by two groups in 2002, has been extensively studied as a major neuronal cold sensor that may also be activated by membrane depolarisation, calcium store depletion, and certain lipids,
as well as by compounds that produce cooling sensations, such as menthol or icilin~\cite{mckemy,peier,voets,vanden,nilius,izqui}. 
The structure of TRPM8 is well defined, and some important binding sites and key domains have been described~\cite{pedretti}. 
Conformational changes underlying the gating of TRPM8 are known mostly for the channels in their closed state~\cite{xu}.
TRPM8 is directly activated by membrane depolarisation, cold, menthol and icilin, as was shown in a single-channel study of the purified TRPM8 protein inserted in artificial lipid bilayers formed from a solution of synthetic 1-palmitoyl-2-oleoyl-glycero-3-phosphocoline (POPC) and 1-palmitoyl-2-oleoyl-glycero-3-phosphoethanolamine (POPE) 
in N-decane~\cite{zak}
However, the lipid membrane environment, G protein activity, intracellular $Ca^{2+}$, and other second messengers are key determinants of channel activity in native cells. One of these key regulators of TRPM8 activity is the minor membrane phospholipid phosphoinositol 4,5-bisphosphate ($PIP_2$), whose concentration is reduced by the activity of the enzyme phospholipase C (PLC), causing a reduction in TRPM8 activity without altering its temperature sensitivity, but shifting its voltage activation range 
to more positive potentials~\cite{yinpark}. 
While $PIP_2$ depletion inhibits TRPM8, activated $G\alpha q$ can also directly inhibit TRPM8 after activation of Gq-coupled receptors without 
the involvement of PLC~\cite{diver}. 
Increases in intracellular calcium concentrations initiate $Ca^{2+}$-dependent desensitization of the TRPM8 channel not only via PLC activation and $PIP_2$ depletion, but also via direct interactions with specific regulatory sites on the TRPM8 protein structure, thus causing channel inactivation during prolonged cooling or 
agonist exposure~\cite{vanden}.
At physiological temperature, e.g. in tissues not exposed to any substantial cooling, a unique biochemical mechanism allows TRPM8 openings following $Ca^{2+}$ store depletion and
activation of $Ca^{2+}$-independent phospholipase A2 (iPLA2)~\cite{fern}.
Thus, we can speculate that some mechanisms modulate the TRPM8's ability to bind agonists, and others modulate gating itself. TRPM8 kinetics at the single-channel
level remain incompletely understood, and the development of simple yet robust methods for quantitatively characterizing TRPM8 transitions between its 
multiple open (two) and closed (five) 
conformational states~\cite{fern} 
is important for research on cold and pain sensation, as well as for drug development.

\section{Materials and Methods}
We have applied theoretical considerations for deriving a  mathematical equation which describes the distribution function for different simple models of TRPM8 functional state with and without correlation. For plotting such distributions and also for computing corresponding distributions based on the simulated data or actual recordings of the activity of heterologously expressed TRPM8 channels, as in this work, we developed a computer program in C language and have used the GCC GNU compiler under Linux, which allows for a fast numerical realization and also utilized the 
GNU Scientific Library (GSL)~\cite{gough2009gnu}.
The main idea of our analysis is to study the difference between the distribution function calculated from the given trace (simulated as in this work or experimentally recorded in the case of a real application for which our approach is targeted) and the predicted distributions for simple models without correlation.

TRPM8 single-channel activity in HEK293 cells expressing the protein was recorded using the cell-attached configuration of the patch-clamp techniques,
as previously described in detail elsewhere~\cite{nicolai}. 
Briefly, the bath solution contained 150 mM KCl, 1 mM MgCl2, 5 mM glucose, and 10 mM HEPES, pH 7.3 (adjusted with NaOH) to bring the resting membrane potential close to 0 mV. The pipettes were filled with solution containing 150 mM NaCl, 1 mM MgCl2, 5 mM glucose, and 10 mM HEPES, pH 7.3 (adjusted with NaOH). To induce cold-dependent activation of TRPM8, the temperature of the bath solution 
was maintained at $\approx 20^\circ C$ (room temperature).
Single-channel currents were filtered at 2 kHz using an eight-pole low-pass Bessel 
filter and sampled at 10 kHz with subsequent cubic spline interpolation in Clampfit 11 software (Molecular Devices, San Jose, CA, USA) 
to decrease the effective sampling interval to 10 $\mu$s. This enabled reliable detection of transitions between the channel`s closed and open states lasting 
longer than 0.16 ms.
\section{Results}

\subsection{Theoretical approach}

In this study, we propose a mathematical approach for the fast analysis of correlation effects in the activity of a single TRPM8 ion channel,
which is generally applicable to other ion channels. As an example of a typical experimental trace, we have plotted a fragment of our experimental record 
in Fig.~\ref{exp} (black line). The same figure also shows a fragment of an idealised trace (the superimposed red line) derived from experimental data. In this theoretical work, we used simulated traces 
from the QUB software~\cite{colquhoun}
(which served as input data for our program) to test our approach. For our theoretical analysis, we have used simulated traces with random noise and more than 10,000 events (i.e., channel transitions between open and closed states).

\myfig{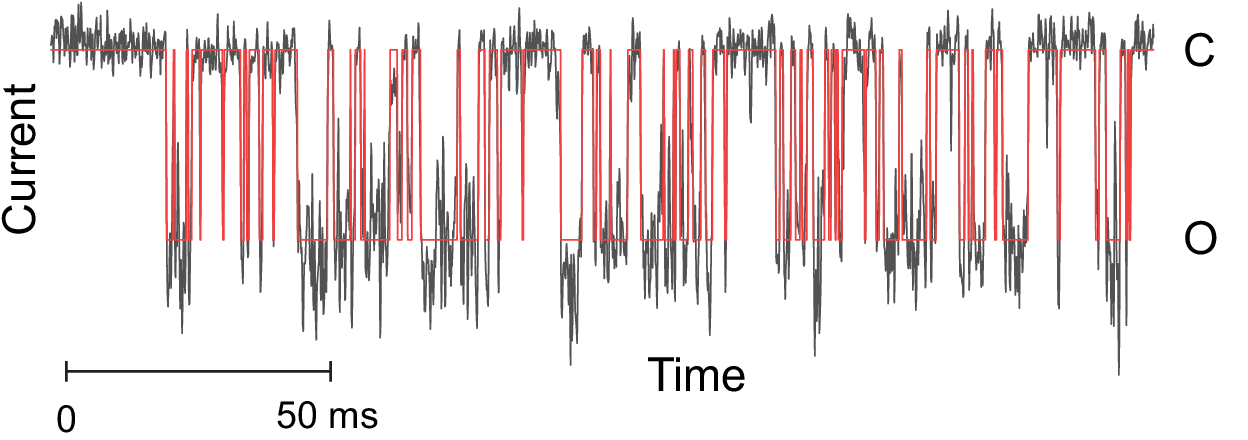}{ Represenattive recording of heterologly expressed in HEK293 cells  TRPM8 single-channel activity recorded at in the cell-attached configuration, hence inward curernt defelections represent out ward currents through trhe membraen patch. The superimposed red line shows idealised curernt trace}{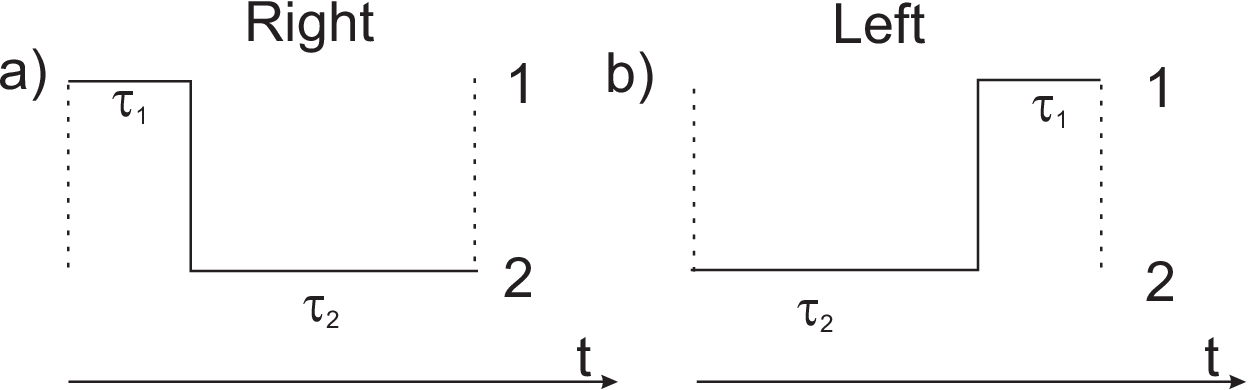}{Scheme for analysis}

Before carrying out the calculation with traces, we have derived formulas for the quantity of interest using some simple models. The first one considers the system (TRPM8 ion channel), which can be in one of the two macrostates (C – close state, O - open state). It is convenient to denote the subsequent duration of 
closed and open states as $\tau_1$ and $\tau_2$ (see Fig.~\ref{Scheme}).


Now, consider the simplest possible situation with only two microstates corresponding to two macrostates. We enumerate them as state 1 (closed C1) 
and 2 (open O1) (Fig.~\ref{Schemes}A).

\mysfig{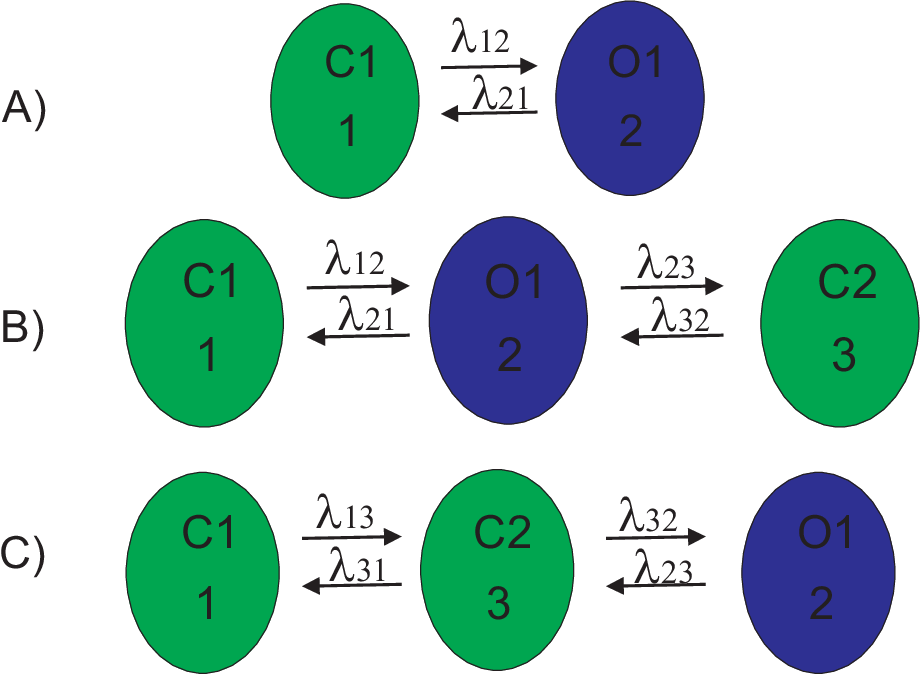}{Simple models for which formulas were deviated}
\mysfigbig{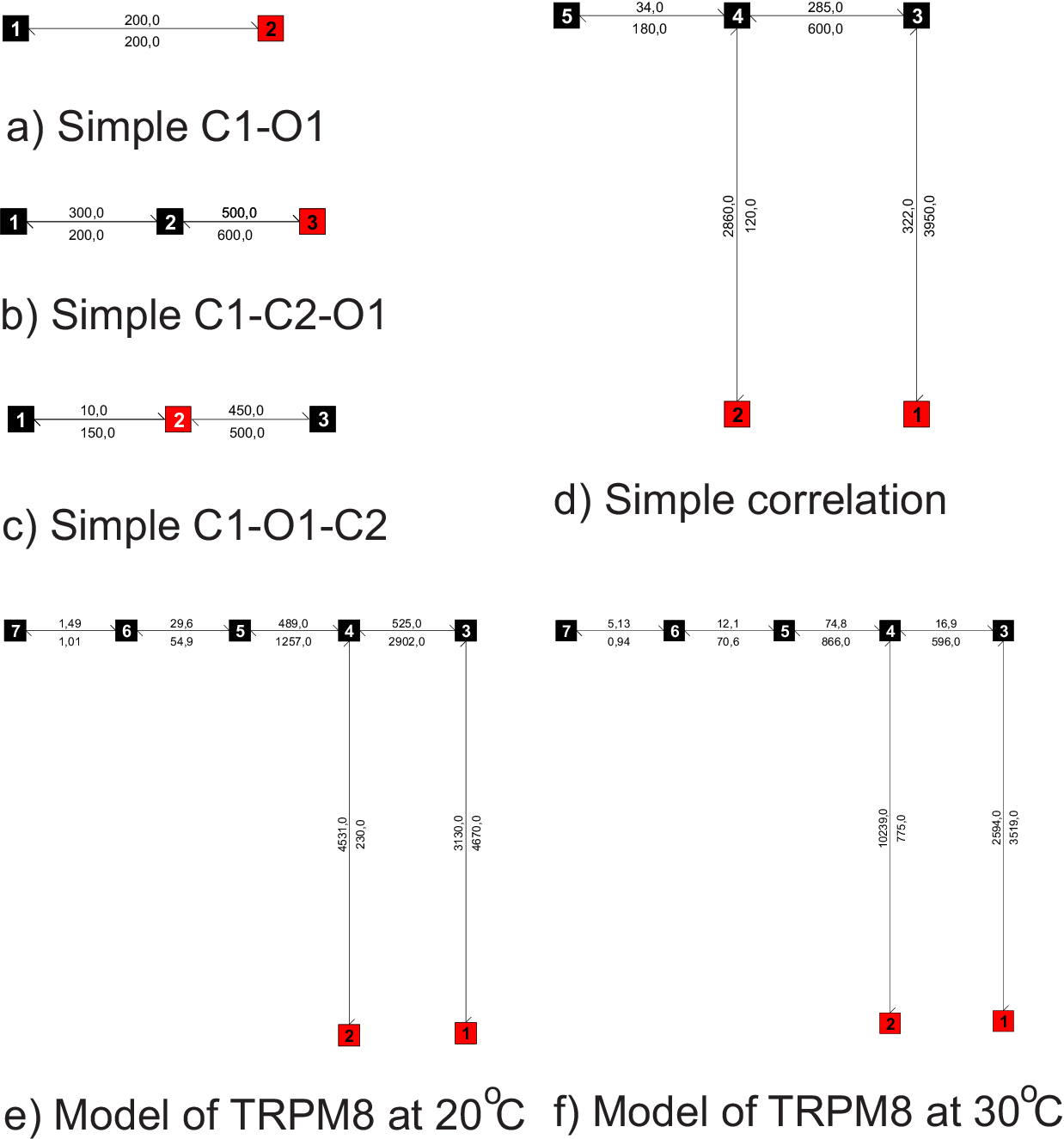}{Models used for calculations. Models e and f are based on our previous single channel study of heterologous TRPM8~\cite{fern}}
The state of the system depends on the parameter t (time) and is also impacted by causality.  Let`s $\xi(t)$ be the state of the system at 
time t and assume that if at some moment s, the system is in the state ``i'', then at some consequent moment - t it could be transferred 
to the state ``j'' with the probability of $p_{ij}(s, t)$ independently of the behaviour of the system before that moment.
Therefore, we will assume that our process $\xi(t)$ is a Markov chain. One may assume that this Markov chain is homogeneous 
$p_{ij}(s,t) = p_{ij}(t - s) (i,j = 1,2)$.
Suppose that at some moment $t_0$ ($t_0 = 0$) we know the state of the system
$\xi(t) = x (x = 1, 2)$. Change of this state will occur at some random moment. In Markov chain theory, the time delay before the system 
changes its state is called the transition waiting time. It is known that this stochastic variable $\tau$ is distributed according to the law:

\begin{eqnarray} \label{eq:eq1}
F_{\tau}(t)=P\{\tau>t|\xi(0)=x\}=e^{-\lambda t}\quad (t>0)\ (x=1,2).
\end{eqnarray}

where $\lambda$  - density of transition from $x$ state.
Let`s denote $\lambda_1$ - density of transition from 1 state
and $\lambda_2$  - density of transition from 2 state. Then, for a simple possible model shown in 
Fig.~\ref{Schemes}A, $\tau_1$ - waiting time for the transition from 1 to 2 and $\tau_2$ -waiting time for the transition from 2 to 1, 
in accordance with equation~(\ref{eq:eq1}):

\begin{eqnarray} \label{eq:eq2}
F_{\tau_1}(t)=P\{\tau_1>t|\xi(0)=1\}=e^{-\lambda_1 t}\quad (t>0),\nonumber \\
F_{\tau_2}(t)=P\{\tau_2>t|\xi(0)=2\}=e^{-\lambda_2 t}\quad (t>0).
\end{eqnarray}

Let us denote as A events in which consequent differences of open and closed state durations are lager then some time t:  $\tau_2-\tau_1>t$

\begin{eqnarray} \label{eq:eq3}
F_A(t)=P\{\tau_2-\tau_1>t|\xi(0)=1,\xi(\tau_1)=2,\xi(\tau_1+\tau_2)=1\} P\{\xi(0)=1\}
\end{eqnarray}

Let us now assume that there is an equilibrium probability that the system is found in state $i$ at some fixed moment,
for example, $t = 0$, and denote it by $\pi_i = P\{\xi(0)=i\}$. By doing this, we assume Ergodicity of the system. 
These probabilities, if they exist, could be calculated from equations

\begin{eqnarray} \label{eq:balance}
\pi_j=\Sigma_{i \in S}\pi_i p_{ij}, j \in S,\nonumber\\
\Sigma_{j \in S}\pi_j=1, j \in S
\end{eqnarray}
where S is the set of possible states, $S=1,2$ in this case. These equations are known as equations of balance.
Now, let us solve them for simple schemes depicted in Fig.~\ref{Schemes}. 

For scheme A transmission matrix $P=(p_{ij})$ is 
$P=\begin{bmatrix} -\lambda_{12} & \lambda_{12} \\
         \lambda_{21} & -\lambda_{21} \end{bmatrix}$
and solving equation of balance, we get

\begin{eqnarray} \label{eq:balancea}
\pi_1=\frac{\lambda_{21}}{\lambda_{12}+\lambda_{21}},\nonumber\\
\pi_2=\frac{\lambda_{12}}{\lambda_{12}+\lambda_{21}}.
\end{eqnarray}

\begin{eqnarray} \label{eq:eq3}
F_A(t)=\pi_1 P\{\tau_2>t+\tau_1|\xi(0)=1,\xi(\tau_1)=2,\xi(\tau_1+\tau_2)=1\}=\nonumber \\
\pi_1 \int_0^{\infty} dx \int_{t+x}^{\infty}f_{\tau_1,\tau_2}(x,y)dy.
\end{eqnarray}

where $f_{\tau_1,\tau_2}(x,y)$ - probability density of joint distribution $\tau_1,\tau_2$,
which in the case of independent events equals to the product of single distribution functions
$f_{\tau_1,\tau_2}(x,y)=f_{\tau_1}(x)f_{\tau_2}(y)$. Due to the definition~(\ref{eq:eq1})
$f_{\tau_1}(x)=-F_{\tau_1}^{\prime}(x)$ and $f_{\tau_2}(x)=-F_{\tau_2}^{\prime}(x)$.

This we can easily evaluate~(\ref{eq:eq3}) for the case of independent events:

\begin{eqnarray} \label{eq:eq4}
F_{A} (t)=\pi_1 \int_0^\infty \mathrm{d}x \int_{t+x}^\infty f_{\tau_1,\tau_2}(x,y)\,\mathrm{d}y = \nonumber \\
\pi_1 \int_0^\infty \mathrm{d}x \int_{t+x}^\infty f_{\tau_1} (x) f_{\tau_2}(y)\,\mathrm{d}y =
\pi_1 \int_0^\infty f_{\tau_1}(x) F_{\tau_2}(t+x)\,\mathrm{d}x
\end{eqnarray}

Let us denote as B events, in which consequent differences of open and closed
state durations are less than -t:  $\tau_2-\tau_1<-t$. Then, in a similar way, we get:

\begin{eqnarray}
F_B(t)=P\{\tau_2-\tau_1<-t|\xi(0)=1,\xi(\tau_1)=2,\xi(\tau_1+\tau_2)=1\} P\{\xi(0)=1\}=\nonumber \\
\pi_1 P\{\tau_1>t+\tau_2|\xi(0)=1,\xi(\tau_1)=2,\xi(\tau_1+\tau_2)=1\}=\nonumber \\
\pi_1 \int_0^{\infty} dy \int_{t+y}^{\infty}f_{\tau_1,\tau_2}(x,y)dx.\nonumber
\end{eqnarray}

And after some simplifications:

\begin{eqnarray} \label{eq:eq5}
F_{B} (t)=\pi_1 \int_0^\infty \mathrm{d}y \int_{t+y}^\infty f_{\tau_1,\tau_2}(x,y)\,\mathrm{d}x = \nonumber \\
\pi_1 \int_0^\infty \mathrm{d}y \int_{t+y}^\infty f_{\tau_1} (x) f_{\tau_2}(y)\,\mathrm{d}x =
\pi_1 \int_0^\infty f_{\tau_2}(y) F_{\tau_1}(t+y)\,\mathrm{d}y
\end{eqnarray}

Using~(\ref{eq:eq2}) from formulas~(\ref{eq:eq4},\ref{eq:eq5}) we get for the simple C1-O1 case:

\begin{eqnarray}
F_A(t)=\pi_1 \frac{\lambda_1}{\lambda_1+\lambda_2} e^{-\lambda_2 t},
F_B(t)=\pi_1 \frac{\lambda_2}{\lambda_1+\lambda_2} e^{-\lambda_1 t}.
\end{eqnarray}

After accounting for equation~(\ref{eq:balancea}) and that for that case $\lambda_1=\lambda_{12},\lambda_2=\lambda_{21}$ we get

\begin{eqnarray}
F_A(t)=\frac{\lambda_{12}\lambda_{21}}{(\lambda_{12}+\lambda_{21})^2} e^{-\lambda_{21} t},
F_B(t)=\frac{\lambda_{21}^2}{(\lambda_{12}+\lambda_{21})^2} e^{-\lambda_{12} t}.
\end{eqnarray}

Obviously there is also possibility that  $-t<\tau_2-\tau_1<t$. Let`s denote a such event as C, then
\begin{eqnarray}
F_C(t)=1-F_A(t)-F_B(t)=\nonumber
1-\frac{\lambda_{12}\lambda_{21}}{(\lambda_{12}+\lambda_{21})^2} e^{-\lambda_{21} t}-\frac{\lambda_{21}^2}{(\lambda_{12}+\lambda_{21})^2} e^{-\lambda_{12} t}.
\end{eqnarray}

We can see that in the absence of correlations, the distributions of $F_{AA}(t)$, $F_{BB}(t)$, and $F_{AB}(t)$ also have a simple form that,
on a logarithmic scale, can be depicted as straight lines. 
Let us consider more complex situations that can be represented by the schemes in Fig. 3B (1C-C2-O1) and Fig. 3C (C1-O1-C2). In Fig. 3B, 
two microstates, 1 and 3, correspond to the C (closed) state, and one microstate, 2, corresponds to the O (open) state. For this scheme

We can see that in the absence of correlations and with only two hidden states (microstates), the distributions $F_A(t)$ and $F_B(t)$ 
have an especially simple form, which, on a logarithmic scale, can be plotted as straight lines. The slope of these lines equals 
$\lambda_1$ and $\lambda_2$. On the same logarithmic scale, these lines also have a difference of their intercepts equal to 
$\ln(\lambda_1/\lambda_2$). Evidently, in the absence of correlation between dwell times of adjacent states, the probability 
function for observing two subsequent ”A” states is the multiplication of single probability functions:

Obviously, that for absence of correlation the probability function for observing
two subsequent ``A'' states is the product of single probability functions:
\begin{eqnarray}
F_{AA}(t)=F_A(t)^2=(\frac{\lambda_1}{\lambda_1+\lambda_2})^2 e^{-2\lambda_2 t},\nonumber \\
F_{BB}(t)=F_A(t)^2=(\frac{\lambda_2}{\lambda_1+\lambda_2})^2 e^{-2\lambda_1 t},\nonumber \\
F_{AB}(t)=F_A(t)F_B(t)=\frac{\lambda_1 \lambda_2}{(\lambda_1+\lambda_2)^2} e^{-(\lambda_1+\lambda_2) t}.\nonumber
\end{eqnarray}
We can see that in case of absence of correlations distribution $F_{AA}(t), F_{BB}(t), F_{AB}(t)$
also have especially simple form which on logarithmic scale can be depicted as straight lines.
Let`s consider more complex situations which can be represented by schemes
Fig.~\ref{Schemes}B (1C-C2-O1) and Fig.~\ref{Schemes}C (C1-O1-C2).
In Fig.~\ref{Schemes} B two microstates 1 and 3 corresponds C (closed) and one microstate 2 corresponds O (open) state.
For the scheme of Fig.~\ref{Schemes}B tarnsition matrix $P=(p_{ij})$ is equals to
$P=\begin{bmatrix} -\lambda_{12} & \lambda_{12} & 0 \\
         \lambda_{21} & -(\lambda_{21}+\lambda_{23})&\lambda_{23}\\
	 0 & \lambda_{32} & -\lambda_{32}
	 \end{bmatrix}$
and solving equation of balance we get
\begin{eqnarray} \label{eq:balanceb}
\pi_1=\frac{\lambda_{21}\lambda_{32}}{\lambda_{12}\lambda_{23}+\lambda_{12}\lambda_{32}+\lambda_{21}\lambda_{32}},\nonumber\\
\pi_2=\frac{\lambda_{12}\lambda_{32}}{\lambda_{12}\lambda_{23}+\lambda_{12}\lambda_{32}+\lambda_{21}\lambda_{32}},\nonumber\\
\pi_3=\frac{\lambda_{12}\lambda_{23}}{\lambda_{12}\lambda_{23}+\lambda_{12}\lambda_{32}+\lambda_{21}\lambda_{32}}.
\end{eqnarray}
Let`s denote $\lambda_1$ - density of transition from 1 state
and $\lambda_2$  - density of transition from 2 state and $\lambda_3$ - density of transition from 3 state.
$\tau_1$ -  waiting time transition from 1 to another state (only 2 in this case see Fig. \ref{Schemes} B and Fig.~\ref{Schemes} C)
and $\tau_2$ - waiting time of transition from 2 to 1 or 3 and  $\tau_3$ - waiting time transition from 3 to 2.

From simple similar derivation like before we have obtained for this C1-O1-C2  Fig.~\ref{Schemes} B case:
\begin{eqnarray}
F_A(t)=(\frac{\pi_1\lambda_1}{\lambda_1+\lambda_2}+\pi_3\frac{\lambda_3}{\lambda_3+\lambda_2} )e^{-\lambda_2 t},\nonumber \\
F_B(t)=\lambda_2(\pi_1\frac{e^{-\lambda_1 t}}{\lambda_1+\lambda_2}+\pi_3\frac{e^{-\lambda_3 t}}{\lambda_3+\lambda_2}) .
\end{eqnarray}
Noticing that for this case $\lambda_1=\lambda_{12},\lambda_2=\lambda_{21}+\lambda_{23},\lambda_3=\lambda_{32}$ and using equation~(\ref{eq:balanceb}) allows
us to explicitly write the previous formula in $\lambda_{ij}$ parameters but we prefer to work with the previous form because it is more compact.

For the last considered model depicted in Fig.~\ref{Schemes}C there is two microstates 1 and 3 which are corresponds to closed state and one microstate 2 corresponds to open state.
For scheme of Fig.~\ref{Schemes} C transmission matrix $P=(p_{ij})$ equals to
$P=\begin{bmatrix} -\lambda_{13} & 0 & \lambda_{13}\\
         0 & -\lambda_{23} &\lambda_{23}\\
	 \lambda_{31} & \lambda_{32} & -(\lambda_{31}+\lambda_{32})
	 \end{bmatrix}$
and solving equation of balance we get
\begin{eqnarray} \label{eq:balancec}
\pi_1=\frac{\lambda_{23}\lambda_{31}}{\lambda_{13}\lambda_{23}+\lambda_{23}\lambda_{31}+\lambda_{13}\lambda_{32}},\nonumber \\
\pi_2=\frac{\lambda_{13}\lambda_{32}}{\lambda_{13}\lambda_{23}+\lambda_{23}\lambda_{31}+\lambda_{13}\lambda_{32}},\nonumber \\
\pi_3=\frac{\lambda_{13}\lambda_{23}}{\lambda_{13}\lambda_{23}+\lambda_{23}\lambda_{31}+\lambda_{13}\lambda_{32}}.
\end{eqnarray}
Making one additional assumption that probability to stay at the state 1 is equal to zero, for model depicted in Fig.~\ref{Schemes} C we get:

\begin{eqnarray} \label{eq:eq14}
F_A(t)= \pi_2 \frac{\lambda_3(\lambda_2-\lambda_1)}{(\lambda_1+\lambda_3)(\lambda_2+\lambda_3)} e^{-\lambda_2 t},\nonumber
F_B(t)= \pi_2 \lambda_3 (\frac{e^{-\lambda_2 t}}{\lambda_2+\lambda_3} -
\frac{e^{-\lambda_1 t}}{\lambda_1+\lambda_3} ) .\nonumber
\end{eqnarray}

Noticing that for that case $\lambda_1=\lambda_{13},\lambda_2=\lambda_{23},\lambda_3=\lambda_{31}+\lambda_{32}$ and using equation~(\ref{eq:balancec}) allows
us to explicitly write the previous formula in $\lambda_{ij}$ parameters but we prefer to work with the previous form because it is more compact.

Summarized formulas acquired for $F_A(t)$ and $F_B(t)$ for considered models are listed in the Table 1. 

\mytab{1}{|c|c|c|c|}{Formulas acquired for $F_A(t)$ and $F_B(t)$ for considered models}{
\hline Panel in Fig.~\ref{Schemes} &
Kinetic scheme & $F_A(t)$ & $F_B(t)$ \\
\hline
A   &   C1-O1	   &  $F_A(t)=\frac{\lambda_1}{\lambda_1+\lambda_2} e^{-\lambda_2 t}$      &   $F_B(t)=\frac{\lambda_2}{\lambda_1+\lambda_2} e^{-\lambda_1 t}$ \\
\hline
B   &	C1-O1-C2   &  $F_A(t)=(\frac{\lambda_1}{\lambda_1+\lambda_2}+\frac{\lambda_3}{\lambda_3+\lambda_2} )e^{-\lambda_2 t}$      &
$F_B(t)=\lambda_2(\frac{e^{-\lambda_1 t}}{\lambda_1+\lambda_2}+\frac{e^{-\lambda_3 t}}{\lambda_3+\lambda_2})$  \\
\hline
C   &	C1-C2-O1   & $F_A(t)=\frac{\lambda_3(\lambda_2-\lambda_1)}{(\lambda_1+\lambda_3)(\lambda_2+\lambda_3)} e^{-\lambda_2 t}$     &
$F_B(t)=  \lambda_3 (\frac{e^{-\lambda_2 t}}{\lambda_2+\lambda_3} - \frac{e^{-\lambda_1 t}}{\lambda_1+\lambda_3} )$ \\
\hline
}

As we can see from Tab. 1, first function $F_A(t)$ has linear dependance in logarithmic scale for all considered models but second function $F_B(t)$ has linear dependance 
only for simple possible case (A). Moreover, as could be demonstrated for more sophisticated models with both numbers of open and closed states greater or equal two, such functions 
($F_A(t)$ and $F_B(t)$) will be nonlinear in logarithmic scale. Therefore, the analysis of distributions we describe here could be indicative for revealing models which contain
 only one open or closed state and could be helpful for identifying a simple possible C1-O1 model for which both distribution functions are linear.

Such distribution functions could be computed and plotted on the base of experimental data which needs only a fast computation that a conventional PC could carry out in less the a second. 
Therefore, proposed approach could be used as express analysis of experimental traces as a first step in formulation of a precise model for a particular type of ion channel. The subsequent section is devoted to detailed description of this computation.

\subsection{Numerical Method and Calculations}

We have calculated the previously introduced distribution function for the simulated QuB traces, which have been saved in the commonly used Axon Text File format. It is widely used for recording experimental ion current traces; therefore, we expect it will simplify the subsequent use of our program for experimental single-channel curernt recordings. For computing such a distribution,
we have used the formula  $F_A^{exp}(t) \approx \frac{N_A(t)}{N}$, where: $N$ is the whole number of open/closed $N$ pairs,
$N_A(t)$ is the number of pairs open/closed states for which the difference in duration of open and close states is 
greater than t.

This is a good approximation because we have always used $N$ greater than thousands. Similarly, $F_B^{exp}(t) \approx \frac{N_B(t)}{N}$,
where: $N$ - whole number of open/closed pairs, $N_B(t)$ - number of pairs open/closed states for which difference in duration of
open and close states where less than $-t$. Second order
distributions were approximated by formula $F_{AA}^{exp}(t)\approx\frac{N_{AA}(t)}{N}$, where:
$N$ - total number of open/closed/open/closed sequences, $N_{AA}(t)$ - number of sequences in which for both
open/closed pairs difference in duration of open and close states was greater than $t$. Similarly,  $F_{BB}^{exp}(t)=\approx\frac{N_{BB}(t)}{N}$,
where: $N$ - total number of open/closed/open/closed sequences, $N_{BB}(t)$ - number of sequences in which for
both subsequent open/closed pairs difference in duration of open and close states was less than $-t$.
And, $F_{AB}^{exp}(t)\approx\frac{N_{AB}(t)}{N}$,
where: $N$ - total number of open/closed/open/closed sequences, $N_{AB}(t)$ - number of
sequences in which for the first open/closed pair, the difference in duration of open and close states is greater 
than $t$, and for the subsequent second open/closed pair, the difference in duration of open and close states
is less than $-t$.
We developed a program capable of idealising the noise in the trace captured during ion current recording (Fig.~\ref{exp}), 
and then computing and plotting the distributions. For testing our program, we performed a simulation in QuB with more 
than a thousand events and with adding noise (to be sure that our program will be capable of working with data really
 close to those of a real experiment) by use of MC models depicted in Fig.~\ref{Schemes} and then we saved 
the obtained traces in ATF format. For simulation modes, a choice has been made that fits well with the single-channel kinetics of TRPM8 ion channels at two different temperatures 
$20^\circ C$ and $30^\circ C$ (see Fig.~\ref{Models}d,e). 

We generated data traces in QuB for all considered models at an acquisition frequency of 20kHz (time step 0.05 milliseconds) and a total of 1 million data points. We have generated noisy data to tailor it to the future experimental recordings, using this approach with real experimental data. 
To calculate the time-dependent distribution function, we used a constant time step of 0.005 s over a 0.2 s interval, yielding 40 data points per trace for the distribution plots. Then we plotted the computed time-dependent distribution function over the interval from 0 to 0.2 seconds.

\mysixfig{nn_Simple}{Simulation for C1-O1: $F_A(t) - red, F_B(t) - black$}{nn_Simple_CCO}{Simulation for C1-C2-O1: $F_A(t) - red,F_B(t) - black$}{nn_Simple_COC}{Simulation for C1-O1-C2: $F_A(t) - red,F_B(t) - black$}{nn_Correlation}{Simulation for simple correlation case: $F_A(t) - red,F_B(t) - black$}{nn_TRPM8_20C}{Simulation for TRPM8-20C channel: $F_A(t) - red,F_B(t) - black$}{nn_TRPM8_30C}{Simulation for TRPM8-30C channel: $F_A(t) - red,F_B(t) - black$}

Fig.~\ref{nn_Simple} and  Fig.~\ref{nn_Simple_CCO} represent curves $F_A(t),F_B(t)$ obtained 
for models depicted in Fig.~\ref{Models}a and Fig.~\ref{Models}b.
Fig.~\ref{nn_Simple_COC} and  Fig.~\ref{nn_Correlation} represent curves $F_A(t),F_B(t)$ obtained 
for models depicted in Fig.~\ref{Models}c and Fig.~\ref{Models}d.
Figs.~\ref{nn_TRPM8_20C} and~\ref{nn_TRPM8_30C} present curves $F_A(t),F_B(t)$ obtained for the models depicted in
Fig.~\ref{Models}e and Fig.~\ref{Models}f.
We have also calculated $F_{AA}(t),F_{AB}(t),F_{BA}(t),F_{BB}(t)$. Figs.~\ref{dd_Simple}--\ref{dd_TRPM8_30C} demonstrate the results 
of these calculations. 

\mysixfig{dd_Simple}{Simulation for C1-O1: $F_{AA}(t) - blue,F_{AB}(t) - green,F_{BA}(t) - red,F_{BB}(t) - black$}{dd_Simple_CCO}{Simulation for C1-C2-O1: $F_{AA}(t) - blue,F_{AB}(t) - green,F_{BA}(t) - red,F_{BB}(t) - black$}{dd_Simple_COC}{Simulation for C1-O1-C2: $F_{AA}(t) - blue,F_{AB}(t) - green,F_{BA}(t) - red,F_{BB}(t) - black$}{dd_Correlation}{Simulation for simple correlation case: $F_{AA}(t) - blue,F_{AB}(t) - green,F_{BA}(t) - red,F_{BB}(t) - black$}{dd_TRPM8_20C}{Simulation for TRPM8 20$^\circ$C channel: $F_{AA}(t) - blue,F_{AB}(t) - green,F_{BA}(t) - red,F_{BB}(t) - black$}{dd_TRPM8_30C}{Simulation for TRPM8 30$^\circ$C channel: $F_{AA}(t) - blue,F_{AB}(t) - green,F_{BA}(t) - red,F_{BB}(t) - black$}

\section{Discussion}

Analysis of single-chamber activity with the aim of establishing its gating mechanism is a daunting process,
which can be very time-consuming owning to the complexity of the interpretation of the stochastic behaviour of the channel gating 
whereby the probability of moving from one state to another depends only on the current state,
not on past history (Markov process)~\cite{fern}. Markov models and rate constants can be calculated from single-channel records,
provided that only one active channel is present in the membrane patch. In the first step,
the numbers of open and closed conformations are determined by assuming that the dwell times in the open and closed states follow exponential
probability distributions. This analysis can be easily performed in Clampfit after running the ``Event detection – Single-Channel Search'' algorithm.
The next step is to determine how the various states are connected, followed by calculating the transition rates (the statistical probability
of a transition between the connected states) over a small time interval. Establishing which states are connected requires performing 
a correlation analysis of the dwell times of adjacent states. In theory, if there is just one transition, which, if interrupted, completely
disconnects open and closed states, no correlation exists between their dwell times (e.g., Fig.~\ref{Models}A,C).
Despite the obvious importance of such correlation analysis, to our knowledge, there is no commercially available software allowing us 
to tackle this problem. Thus, when establishing the kinetic gating mechanism of TRPM8 we used a combination of analysis in Clampfit 
to obtain idealized traces, QUB software, custom scripts in Origin, and custom software written in
K.L. Magleby`s laboratory (Magleby \& Song, 1992) and MATLAB (The MathWorks) in order to comprehensively characterise the correlations 
both in 2D and 3D formats (Magleby \& Song, 1992). To tackle this problem, we have now developed a fast computational algorithm 
that enables reliable visualization of correlations between the dwell times of adjacent open and closed states of an ion channel, if any exist.
As we can see from the analysis of time dependence of calculated distribution curves (Figs.~\ref{nn_Simple}-\ref{nn_TRPM8_30C}) on logarithmic timeline, both curves are linear only for the simplest possible model C1-O1 depicted in Fig.~\ref{nn_Simple}. This is consistent with our prediction on the base of analytical formulas summarized in Table 1: the only model which has only one exponential function for both distribution functions (for both columns in Table 1) is for model A which corresponds to the C1-O1 gating mechanism. For other cases, the first distribution is linear (red line), but the second (black) is non-linear, which also agrees with our prediction 
on the basis of analytical formulas, for the considered C1-C2-O1 and C1-O1-C2 models (B and C models in Table 1), only the first distribution ($F_A(t)$) has one exponent in their formula; therefore, only this one should be linear on the logarithmic timescale. If we swap closed with open states then the distribution $F_A(t)$ swaps with FB(t). This allows us to predict that, for the cases O1-O2-C1 and O1-C1-O2, $F_A(t)$ will be nonlinear on a logarithmic timescale, while FB(t) should be linear. Therefore, based on the analysis of such a distribution of experimental traces, we could propose reasons for building a Markov model. If both distribution curves are linear, we could approximate with the C-O model. If $F_A(t)$ is linear but $F_B(t)$ is not, we should choose between C1-C2-O1 or C1-O1-C2 schemes. If $F_B(t)$ is linear but
$F_A(t)$ is not, so we should choose between O1-O2-C1 or O1-C1-O2 schemes. Of course, for accounting for the kinetics of ion channels, additional hidden states could be added based on standard analysis. A similar tendency has also been seen for the second-order curves of Figs.~\ref{dd_Simple},~\ref{dd_Correlation},~\ref{dd_Simple_CCO},~\ref{dd_TRPM8_20C},~\ref{dd_Simple_COC},~\ref{dd_TRPM8_30C}.
Therefore, deviations of calculated dependences from linearity on a logarithmic timescale allow us to predict the presence of correlations in channel kinetics. Namely, the Pearson correlation coefficient can be calculated, and its absolute value can be (statistically) compared with 1.0.

As an additional step of statistical analysis, linear and quadratic curve fitting could be performed, 
and the R-square value for different regression models (linear and quadratic) could be calculated. 
As is well known, R-squared characterises the percentage of the variance in the deviations between the data and the 
corresponding model values that the model is capable of explaining. Therefore, if Pearson’s correlation is small 
enough and if the R-square value is higher for the quadratic model than for the linear model, then it is evidence of 
linearity of the dependence. Of course, the randomness of residues should also be checked for both models
(linear and quadratic).

\section{Summary}

In this study, we propose an efficient computational technique applied to noisy simulated data that models the kinetics of the TRPM8 ion channel. The main advantage of the developed approach is that it can be applied to real experimental traces, probing deviations of the corresponding plots from simple linear dependence to investigate ion channel kinetics and identify correlations between dwell times of adjacent closed and open states of an ion channel.

\bibliographystyle{apsrev}
\bibliography{Ogloblya}

\end{document}